\documentclass[prl,aps,twocolumn,preprintnumbers]{revtex4}

\usepackage{graphicx}
\usepackage{dcolumn}
\usepackage{bm}

\begin{document}

\preprint{PRL {\bf 94}, 048901(2005).}

\title{Comment on ``Secure Communication using mesoscopic coherent states", Barbosa et al, Phys Rev Lett 90, 227901
(2003)}

\author{Z. L. Yuan and A. J. Shields}
\affiliation{Toshiba Research Europe Limited, 260 Cambridge Science
Park, Cambridge CB4 0WE, UK}

\date{18 August 2003}

\pacs{03.67.Dd, 42.50.Ar, 42.50.Lc}

\maketitle

In a recent letter, Barbosa {\it et al.}\cite {Barbosa} claim that
secure communication is possible with bright coherent pulses, by
using quantum noise to hide the data from an eavesdropper. We show
here that the secrecy in the scheme of Barbosa et al is unrelated to
quantum noise, but rather derives from the secret key that sender
and receiver share beforehand.

In Ref.\cite{Barbosa} binary data is encoded upon $M/2$
non-orthogonal bases, chosen using a key $K'$, which is expanded
from a short, shared seed key $K$ using a stream cipher.  In the
example given, each bit is encoded as one of the $M/2$ possible
different linear polarization bases $\phi_l= l/M$, where
$l(=0,\cdot\cdot\cdot,M/2-1)$ is defined by $log_2(M/2)$ bits of
$K'$. For $l$=even, bit=0 is represented by $\phi_l= \pi l/M$ and
bit=1 by $\phi_l=\pi l/M + \pi/2$, while for $l=$odd, bit$=0$ is
represented by $\phi_l= \pi l/M + \pi /2$ and bit$=1$ by
$\phi_l=\pi l/M$. Since the bit values of adjacent bases are encoded
with opposite parity, Eve will be unable to recover the data by
direct measurement due to the Shot noise.  On the other hand, since
Bob has access to $K'$, he can rotate the measurement basis to the
appropriate angle and measure the parity to determine the bit value.
A similar scheme has been suggested in Ref. \cite{Hirota} using
intensity modulation.

We show that the security of this scheme is very closely related to
that of the one time pad.  To realize this, notice that it is not
necessary for Bob to apply the key $K'$ prior to his measurements.
Bob could gain exactly the same information ($I_{AB}$) by firstly
performing a measurement of the polarization angle and then using
$K'$ to determine the bit value from this measurement.  It is clear
from this line of argument, that Eve can make identical measurements
to Bob and will obtain identical information ($I_{AE}=I_{AB}$) to
him from this measurement.  Alice and Bob will therefore be unable
to expand the secret information they share, since the rate of this
expansion is given by $\Delta I=I_{AB}-I_{AE}=0$.\cite {Mauer}
Unlike Bob, Eve will be unable to interpret these measurements as
she is not in possession of $K'$. Clearly, the secrecy of the data
relies entirely upon the secrecy of the key. Quantum noise does not
play any role, since Eve (like Bob) needs only determine the parity
of each pulse and not its exact polarization angle.

It is well known that the one-time pad is secure from
crypto-analysis provided the key material is used only once.  The
security is compromised if the key is used repeatedly.  This
requirement of fresh key material renders the one-time pad
impractical for most applications. The scheme of Ref.\cite{Barbosa}
would be very attractive, if quantum noise could allow an expanded
key to be used securely with the one-time pad.  Unfortunately,
however, this is not the case.  If Alice and Bob use an expanded
key, Eve can analyze her measurements of the polarization angles of
each pulse to determine the seed key and data.  This is readily
apparent from considering the following, very simple, eavesdropping
strategy.

Eve measures the linear polarization of each pulse. If she measures
an angle 0$\le$$\varphi$$<$$\pi$/2, she assigns the bit$=0$, while
if she determines $\pi/2\le\varphi<\pi$ she assigns bit=1.
 We can regard the bit sequence that Eve determines in this way,
 as the encrypted bit sequence {\bf E}.  Because Alice alternates the parity
  of the encoded bits between adjacent bases, 50$\%$ of the bits in the encrypted sequence {\bf E}
will agree with the corresponding bit in the original data D and
$50\%$ will differ.  Thus Alice's and Eve's data can be related by
{\bf D$\bigoplus$L=E}, where {\bf L}=0 if Alice chooses {\it l}=even
and {\bf L}=1 if {\it l}=odd. To recover the original data Eve must
determine the sequence L. Notice that the simple relationship
connecting the original and encrypted data is exactly that of the
one time pad and, as for the one-time pad, key expansion will render
the scheme vulnerable to crypto-analysis.

Using this eavesdropping strategy Eve will obtain some errors in her
encrypted data sequence when the polarization angle lies very close
to 0 or $\pi /2$.  We stress that this is simply because of the very
simple decoding strategy that we have adopted for Eve.  The rate of
these errors will decrease as the intensity of the Alice's pulses
increases, illustrating that bright pulses degrade the security.
Indeed optimal uncertainty is recovered for the case that Alice
sends single photon pulses.

In summary, the secrecy of the scheme in Ref. \cite{Barbosa} does
not rely upon quantum noise, but rather is closely related to that
of the one-time pad. Furthermore, the requirement to use
$log_2(M/2)$ key bits for each data bit, also renders the scheme
considerably less efficient than the one time pad, for which just
one key bit is needed per data bit. Nevertheless, it is an
alternative data encryption scheme, which, though not perfectly
secure and less efficient than the one-time pad, may find
applications in optical communication technology.


\begin{thebibliography} {99}
\bibitem{Barbosa}
G.~A.~Barbosa {\it et al}, Phys.\ Rev.\ Lett.\ {\bf 90},
227901(2003).
\bibitem {Hirota}
O.~Hirota {\it et al.}, quant-ph/0212050(2002).
\bibitem {Mauer}
U.~Mauer, IEEE Trans. Inf. Theory {\bf 39}, 733(1993).


\end{thebibliography}
\end{document}